 \documentclass[sigconf]{acmart}

\usepackage{placeins}
\AtBeginDocument{%
  }

\setcopyright{none}
\acmYear{2026}
\copyrightyear{2026}
\acmDOI{}
\acmISBN{}

\acmConference[MoDeVVa '26]{23rd Workshop on Model Driven Engineering, Verification and Validation}{October 04--06, 2026}{M{\'a}laga, Spain}

\setcopyright{none}
\acmDOI{}
\acmISBN{}

\begin{document}

\title{LLM Ensemble Fault Classification for Automotive HiL Validation}

\author{Hamza Ouarrad}
\email{hamza.ouarrad@tu-clausthal.de}
\correspondingauthor
\authornotemark[1]
\affiliation{%
  \institution{Institute for Software and Systems Engineering}
  \city{Clausthal-Zellerfeld}
  \country{Germany}
}

\author{Mohammad Abboush}
\email{mohammad.abboush@tu-clausthal.de}
\affiliation{%
  \institution{Institute for Software and Systems Engineering}
  \city{Clausthal-Zellerfeld}
  \country{Germany}}

\author{Andreas Rausch}
\email{andreas.rausch@tu-clausthal.de}
\affiliation{%
  \institution{Institute for Software and Systems Engineering}
  \city{Clausthal-Zellerfeld}
  \country{Germany}
}

\renewcommand{\shortauthors}{Ouarrad et al.}

\begin{abstract}
Automotive HiL validation generates large multivariate test recordings whose
analysis remains challenging due to manual review effort, rule-based
limitations, and the need for explainable diagnostic decisions. Recent
machine-learning and deep-learning approaches have improved fault diagnosis,
but they often require large labelled datasets, generalise poorly across
operating conditions, and provide limited insight into their predictions. This
paper proposes an explainable multi-LLM ensemble framework for sensor-level
fault classification in automotive validation. The framework uses compact
evidence representations of fault-injection recordings and combines the outputs
of heterogeneous large language models to improve diagnostic robustness,
ranking quality, confidence reliability, and interpretability. The approach is
evaluated on gasoline-engine and electric-vehicle HiL systems across three
driving settings and ten single-fault classes. Among the individual models,
Mistral Small~24B provides the strongest overall single-model trade-off,
achieving 0.903 Top-1 accuracy, 0.887 MCC, and the lowest Brier score of
0.102. The final Top-3 ensemble combines Mistral Small~24B, Qwen2.5~32B, and
Phi-4~14B using confidence-weighted voting, improving the scenario-averaged
results to 0.917 Top-1 accuracy, 0.913 macro F1, and 0.902 MCC, while also
providing the best calibration among the tested ensemble strategies. A Top-5
ensemble does not improve over the Top-3 configuration, indicating that model
complementarity is more important than ensemble size. The results show that
coordinated multi-LLM reasoning can support robust, calibrated, and
engineer-interpretable fault classification for automotive HiL validation.
\end{abstract}


\maketitle

\section{Introduction}
In the automotive domain, the development of safety-critical software systems is governed by the ISO 26262 functional safety standard \cite{url}, which prescribes rigorous verification and validation across the system development life cycle. These activities span successive integration levels of the V-model and are collectively termed X-in-the-Loop (XiL) testing \cite{tibba2016testing}. In the final validation phase preceding series production, real test drives on public roads, complemented by hardware-in-the-loop (HIL) simulations integrating actual electronic control units (ECUs) and physical components, validate the system against its functional safety requirements under representative operating conditions \cite{klitzke2019real,abboush2024virtual}.

Such campaigns generate large-volume test recordings that manifest as multivariate time series data with nonlinear dynamics and multi-mode behaviour \cite{theissler2013anomaly}. A single advanced driver-assistance (ADAS) test run may record hundreds of signals at several kilohertz over hours, yielding gigabytes per run, while a full release campaign may comprise thousands of runs \cite{lee2017collecting}. Current industrial practice relies predominantly on tool-based analysis with predefined thresholds and pattern-matching rules \cite{theissler2013detecting}. While effective for flagging anomalous patterns, such techniques are limited to anomaly detection and offer little support for identifying fault type and severity \cite{goina2024enhanced}. Moreover, the governing rules must be authored manually, demand expert maintenance, and cannot detect unanticipated failure modes. Manual review at this scale is therefore cognitively prohibitive, creating a bottleneck that limits validation throughput. There is thus a pressing need for intelligent diagnostic frameworks that integrate data-driven evidence and engineering knowledge to support explainable fault detection and diagnosis in HIL validation, consistent with ISO 26262 \cite{url}.

Recently, data-driven fault detection and diagnosis (FDD) approaches have been introduced to reduce the reliance of conventional model-based and knowledge-based methods on accurate system models and expert knowledge. With growing large-scale time-series data, machine learning (ML) and deep learning (DL) techniques have shown strong automotive fault classification performance \cite{dai2013knowledge}, including for electric vehicle powertrains. Nevertheless, these methods display restricted generalisability to unseen operating conditions and configurations \cite{baccari2024anomaly}; most supervised DL models assume stationary distributions and require retraining from scratch when new faults emerge, yielding low training efficiency and limited scalability in safety-critical settings \cite{zhang2025survey}. They also demand large labelled datasets across diverse fault scenarios and remain largely opaque, with domain expertise still integrated through manual expert involvement \cite{lei2025research}.

In parallel, large language models (LLMs) such as GPT, LLaMA, Qwen, and Phi have shown remarkable capabilities in logical reasoning and heterogeneous information processing \cite{zheng2024empirical}. Recent studies apply them to fault diagnosis by converting time-series data into textual representations and exploiting language-based reasoning, and their self-supervised cross-domain pre-training enables efficient fine-tuning for downstream tasks \cite{lin2025fd}. However, most current LLM-based frameworks rely on a single model, prompt template, or reasoning path. Such standalone settings are susceptible to prompt sensitivity, inconsistent outputs, overconfident errors, and reduced robustness when evidence is incomplete or ambiguous, which is especially consequential for safety-related applications where diagnostic stability and traceability matter as much as accuracy. By contrast, ensemble learning has long improved generalisation, reduced variance, and exploited complementary model strengths in ML and fault diagnosis, yet this principle has not been systematically developed for sensor-level vehicle fault detection and diagnosis using multiple LLMs as coordinated diagnostic agents.

To address this gap, this article proposes a novel ensemble LLM framework for sensor-level fault detection and diagnosis from real and digital test-drive vehicle signals. The central idea is to transform windowed multivariate sensor data into structured prompts summarising local signal behaviour, operating context, and candidate fault evidence, then query several heterogeneous LLMs using zero-shot and few-shot prompting. Each model independently produces a detection decision, a fault-type prediction, a confidence estimate, and a human-readable explanation, which are combined through an ensemble fusion mechanism designed to improve robustness, mitigate single-model bias, and yield more reliable decisions than any standalone LLM. The framework is evaluated on data from both gasoline-engine and electric vehicle systems, generated using high-fidelity simulation models executed in real time, reflecting the complexity of industrial validation practice. To the best of the authors' knowledge, this is the first study to develop a coordinated multi-LLM ensemble for sensor-level fault detection and diagnosis in automotive systems.

The main contributions of this work are as follows:
\begin{itemize}
    \item A first ensemble LLM-based framework for sensor-level fault detection and diagnosis in vehicle systems, validated on real and digital test-drive data from gasoline-engine and electric vehicle platforms.
    \item A prompt engineering strategy for transforming multivariate time-series sensor windows into zero-shot and few-shot diagnostic prompts suitable for LLM reasoning.
    \item A multi-LLM decision fusion mechanism that combines fault labels, confidence levels, and explanations to overcome the limitations of standalone LLM diagnosis.
    \item A comparative evaluation against single-LLM approaches across multiple fault scenarios.
\end{itemize}
The remainder of this article is organised as follows. Section 2 reviews the relevant literature. Section 3 presents the proposed approach. Section 4 describes the implementation and case study. Section 5 reports and discusses the results. Section 6 concludes the article and outlines future directions.

\section{Related work}

Automotive FDD has progressed from shallow classifiers towards deep architectures that learn discriminative representations directly from raw multivariate signals. Convolutional–dense pipelines have been applied to sensor-fault detection and prognostic health-index forecasting in autonomous-driving stacks, with Safavi et al. \cite{safavi2021multi} reaching 99.84\% detection accuracy on the Audi A2D2 dataset but restricting evaluation to four canonical fault classes injected via a static distribution model, while recurrent architectures have dominated diagnostics in electrified powertrains, exemplified by the LSTM model of Kaplan et al. \cite{kaplan2021fault}, which cut MAPE from 10.13\% to 2.06\% over a shallow ANN, albeit with fault injection performed only at the Simulink level. Hybrid and ensemble formulations have broadened coverage: a two-class/one-class ensemble attaining a 77\% F2-score on OBD-II recordings \cite{theissler2017detecting}, a multi-label LSTM–Random Forest ensemble with a GRU denoising autoencoder reaching 99.43\% accuracy on concurrent faults under HIL-generated noise \cite{abboush2023intelligent}, and a Random Forest classifier coupled with fault-tolerant control on a TruckMaker-based HIL platform \cite{raveendran2020brake}; reviews further note that many algorithms simplify actuator dynamics and neglect parameter drift \cite{pietrowski2024fault}. Collectively, these methods improve accuracy yet remain confined to narrow fault catalogues, single-fault scenarios and single-platform validation, depend on large labelled datasets, and offer limited interpretability — constraints that motivate both ensemble strategies for robustness and language-based reasoning for knowledge integration. Ensemble learning mitigates the variance and bias of single estimators by combining complementary base learners through stacking, blending and weighted fusion. In rotating-machinery diagnosis, a deep blending ensemble of CNN base learners with an ECOC meta-learner attained 98.9\% accuracy and sharply reduced missed detections for weak bearing faults at low SNR \cite{inyang2023diagnosis}. Stacked heterogeneous ensembles perform comparably in power systems — an RF–LSTM-tuned-KNN stack optimised by PSO reached 99.96\% on transmission-line faults \cite{anwar2025robust}, and a DNN/LSTM/Bi-LSTM stack with a logistic-regression meta-learner classified photovoltaic faults at 98.62\% (clean) and 94.87\% (noisy) \cite{lodhi2023novel}. Weighted and probability-based fusion further balances diverse architectures, as in a CNN–BiLSTM–BiGRU weighted ensemble reaching 99.62\% on UAV faults under wind disturbance \cite{huang2024unmanned} and a weighted-probability ensemble of pretrained CNNs diagnosing induction-motor faults at up to 99.60\% \cite{ali2025improved}; classical RF/XGBoost on wavelet features exceed 99\% on motor-current data \cite{nishat2020bearing}. Transferability and efficiency are addressed by a Bayesian-optimised 1D-CNN ensemble for cross-machine transfer with few labels \cite{pacheco2022deep} and the parallel ML–DL stack EMDL-FDD, reaching 99.79\% while cutting computation time by up to 86\% \cite{al2023parallel}. Across these studies ensembles consistently outperform their constituent models and resist noise, but two limitations recur: the computational and memory overhead of combining multiple learners constrains real-time deployment, and all operate over numerical features and yield categorical decisions without interpretable justification. Crucially, the ensemble principle has not been transferred to language-based reasoning, where model diversity could mitigate instability while simultaneously generating explanations.

The reasoning and knowledge-integration strengths of LLMs have prompted growing interest in automotive diagnosis. One textual strand classifies large fleets of fault-symptom claims \cite{pavlopoulos2024automotive} and produces evidence-grounded reports via CTGAN–RAG–GPT-4o pipelines \cite{mahale2025automated}, attaining high readability but only moderate factual accuracy. A retrieval strand embeds structured knowledge into diagnostic assistants, including RAG combined with knowledge graphs for New Energy Vehicles \cite{zhang2025large}, distributed on-vehicle diagnosis under network constraints \cite{chen2026personalized}, and HIL-GPT, whose domain-adapted compact models surpass larger counterparts in HIL test-sequence retrieval \cite{feng2025smarter}. Most directly, an evaluation of zero-shot, Chain-of-Thought, and fine-tuned LLMs on Connected-Vehicle Basic Safety Messages reports that base models lack intrinsic temporal reasoning, perform poorly on trend and stuck-at faults, and show run-to-run instability \cite{das2025llm}. Related applications span autonomous-driving violation diagnosis \cite{lu2024diavio}, API testing \cite{wang2025automating}, and hazard-analysis assistance \cite{diemert2023can}, with performance consistently declining as system complexity increases. Taken together, automotive LLM research concentrates on textual and retrieval tasks, with fault detection from raw signals still at an early stage, and its recurring weaknesses, namely limited temporal reasoning, hallucination and overconfidence, prompt sensitivity with inconsistent single-model outputs, and scarce automotive validation, correspond precisely to the failure modes that ensembles mitigate in conventional FDD.

Yet the ensemble principle has not been developed for multi-LLM diagnosis, and no study coordinates several heterogeneous LLMs as complementary diagnostic agents for sensor-level fault detection. This motivates the present work, which converts windowed multivariate sensor data into structured zero-shot and few-shot prompts, queries several heterogeneous LLMs each returning a decision, fault type, confidence and explanation, and fuses these outputs to deliver more robust and interpretable diagnoses than any standalone model, validated on gasoline-engine and electric vehicle systems under high-fidelity real-time simulation.

\section{Methodology}

\begin{figure*}[t]
    \centering
    \includegraphics[
        width=\textwidth,
        height=0.78\textheight,
        keepaspectratio
    ]{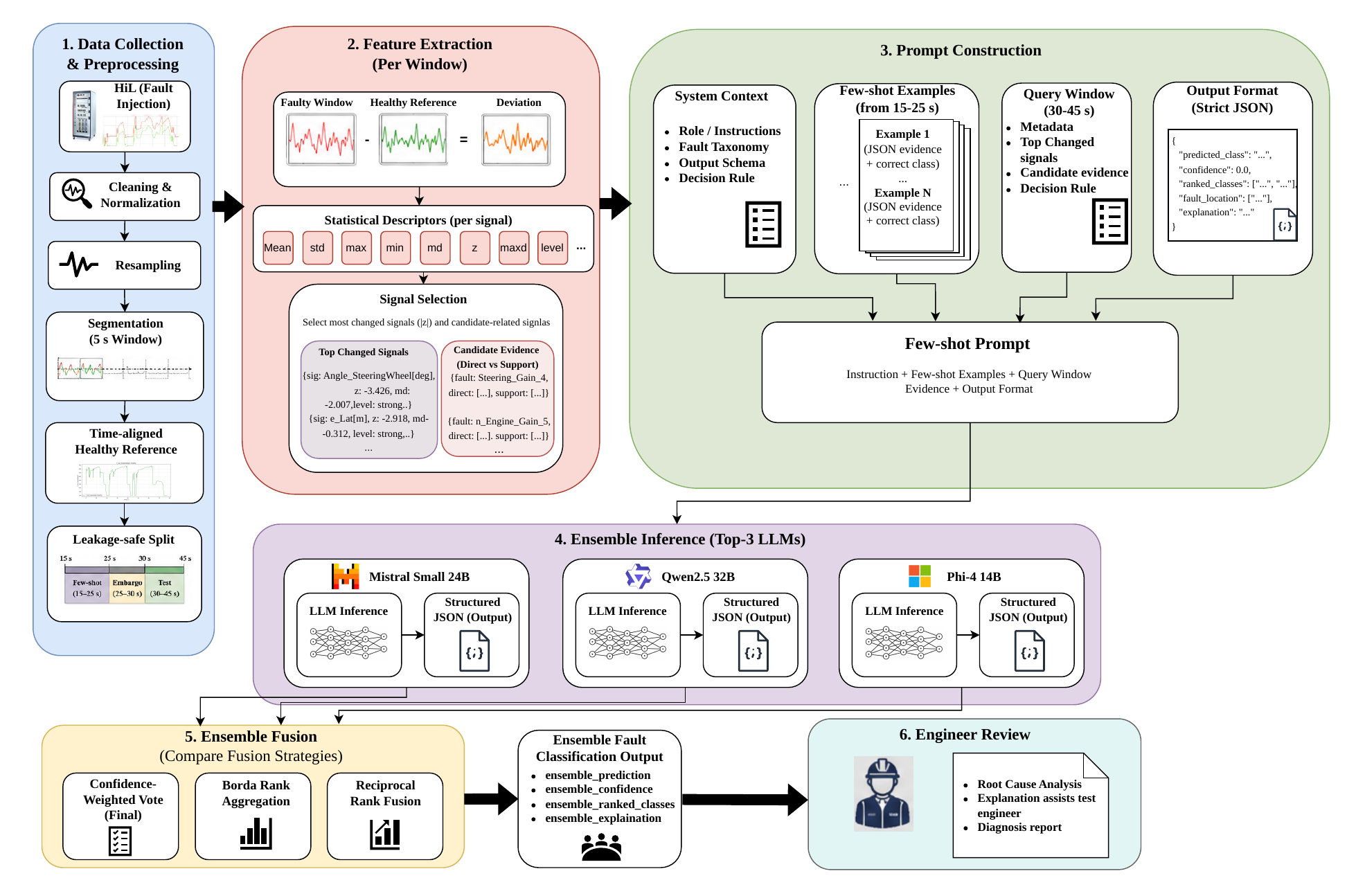}
    \caption{Overview of the proposed ensemble-based fault classification framework.}
    \Description{Workflow diagram showing preprocessing, compact evidence extraction,
    prompt construction, ensemble inference, ensemble fusion, and engineer review.}
    \label{fig:framework_overview}
\end{figure*}

The proposed methodology is shown in Fig.~\ref{fig:framework_overview}. It transforms HiL fault-injection recordings into compact diagnostic evidence, submits the same structured prompt to several selected LLMs, and combines their outputs through ensemble fusion. The objective is not to classify raw time-series values directly, but to compare faulty behaviour with a time-aligned healthy reference and use the resulting deviation evidence for fault classification. 

\subsection{Data Preprocessing and Windowing}
\label{subsec:preprocessing_windowing}

For each driving scenario, the corresponding healthy run is used as a
time-aligned reference trajectory. This is necessary because the measured
signals are influenced by the driving profile itself. Therefore, the framework
focuses on deviations from healthy behaviour in the same time interval rather
than on absolute signal values.

The analysis is restricted to the active fault interval between 15~s and
45~s. This interval excludes the initial transition before the fault becomes
stable and the later fault-removal phase. The time series is segmented into
5~s windows with a step size of 1~s. For every faulty window, the same time
interval is extracted from the healthy reference run, enabling a direct
fault-versus-healthy comparison.

To avoid temporal leakage, the windowed data are divided into separated
time-based regions. Windows from 15--25~s are used to construct labelled
few-shot examples, windows from 25--30~s are used as an embargo region, and
windows from 30--45~s are used for final evaluation. This prevents the examples
provided to the LLM from overlapping temporally with the test windows.

\subsection{Compact Evidence Extraction}
\label{subsec:evidence_extraction}

Each faulty window is compared with the corresponding healthy window from the
same time interval. For every signal, compact statistical descriptors are
computed, including mean difference, mean absolute deviation, maximum
deviation, zero and saturation fractions, and a standardized deviation score.
The standardized deviation expresses how strongly a signal differs from the
healthy reference relative to normal variation.

The prompt does not contain raw time-series samples. Instead, each window is
summarized using compact evidence objects. These include the most changed
signals and candidate-specific evidence. Direct root-fault evidence is
separated from propagated support evidence, so that the model is encouraged to
identify the fault source rather than only downstream signal effects. This
keeps the input concise while preserving the diagnostic relation between a
fault location and its observable behaviour.

\subsection{Few-Shot Prompt Construction}
\label{subsec:prompt_construction}

The prompt defines the LLM as an automotive HiL fault-diagnosis assistant and
formulates the task as fault-only fault classification. The healthy recording is used only as a time aligned reference and is not included as a valid output class. Each prompt contains the system context, the candidate fault classes, a small set of leakage-safe few-shot examples from the 15--25~s interval, and one unlabeled query window from the 30--45~s test interval.

The query window is represented by compact statistical evidence rather than raw time-series samples. The prompt separates direct root-cause evidence from propagated support evidence, encouraging the model to identify the fault source rather than only downstream responses. This structure is intended
to make the LLM decision more diagnostic and less dependent on secondary signal changes.

All models are instructed to return a structured JSON object. This enables automatic parsing of the predicted class, confidence value, ranked alternatives, fault location, and explanation. The required output format is:

\begin{verbatim}
{
  "predicted_class": "one candidate class",
  "confidence": 0.0,
  "ranked_classes": ["best", "second"],
  "fault_locations": ["location"],
  "explanation": "short root-evidence explanation"
}
\end{verbatim}

The predicted class is used for Top-1 evaluation, while the ranked list is used
for Top-2 evaluation. The confidence score supports calibration analysis, and
the explanation gives the test engineer a concise justification of the model
decision.

\subsection{Top-3 LLM Ensemble Inference and Fusion}
\label{subsec:ensemble_inference}

The same few-shot prompt is submitted independently to the three strongest
single models: Mistral Small~24B, Qwen2.5~32B, and Phi-4~14B. Each model
returns a structured prediction with a confidence value, ranked candidates, and
an explanation. The individual outputs are then combined using three fusion
strategies: confidence-weighted voting, Borda rank aggregation, and reciprocal
rank fusion.

Confidence-weighted voting uses the predicted class and confidence score of
each model to favour predictions supported by stronger model confidence. Borda
rank aggregation combines the ranked candidate lists by assigning higher
scores to candidates placed higher by the individual models. Reciprocal rank
fusion similarly uses the candidate rankings, but gives stronger weight to
classes that appear near the top of a model's ranked output. The final ensemble
output contains the selected fault class, an aggregated confidence score,
ranked alternatives, and the supporting explanation for engineering review.

\section{Case Study and Implementation}

\label{sec:case_study}

\subsection{ASM Gasoline-Engine HiL System}

\label{subsec:asm_hil}

The first case study uses a dSPACE ASM gasoline-engine system operated in a closed-loop HiL environment. The setup includes coupled driver, vehicle, and engine signals, so that injected faults do not remain isolated but propagate through physically related variables. This makes the case study suitable for evaluating whether the proposed LLM ensemble can distinguish direct root-cause evidence from secondary effects. The recorded signals cover driver inputs such as accelerator, brake, and steering, as well as engine and vehicle responses such as engine speed, throttle position, torque, and vehicle velocity.

\subsection{Electric-Vehicle HiL System}

\label{subsec:ev_hil}

The second case study considers an electric-vehicle powertrain system in the same HiL validation context. Compared with the gasoline-engine system, this setup focuses on electrical propulsion and battery-related dynamics. The recorded signals include high-voltage battery measurements, DC-link variables, rear electric-machine speed and torque, steering torque, and vehicle speed. Together, the gasoline-engine and electric-vehicle platforms allow the framework to be evaluated across two different propulsion architectures and signal-interaction structures.

\subsection{Dataset Description}
\label{subsec:dataset_description}

The dataset was generated from automated fault-injection experiments in the two
HiL configurations. Each recording corresponds to one experimental run under
healthy or faulty conditions and contains approximately 7{,}700 samples with a
fixed sampling interval of 0.01~s. The gasoline-engine recordings contain 15
signal columns, while the electric-vehicle recordings contain 17 signal
columns. For each faulty run, the corresponding healthy run of the same driving
profile is used as a time-aligned reference trajectory. The evaluation covers
three settings and ten single-fault classes. The gasoline-engine highway
setting includes accelerator-pedal, brake-pedal, engine-speed, steering, and
throttle faults. The electric-vehicle highway setting contains high-voltage
battery, rear electric machine speed, and steering faults. The gasoline-engine
urban setting includes accelerator-pedal-gain and engine-speed-noise faults.
This combination allows the method to be tested on different systems, driving
conditions, and fault signatures. The experiments evaluate both zero-shot and
few-shot prompting. The zero-shot setting uses only the query-window evidence,
whereas the few-shot setting adds leakage-safe labelled examples and the
compact direct-evidence representation described in
Section~\ref{subsec:prompt_construction}.

\section{Results and Discussion}
\label{sec:results}

The results are reported as scenario-averaged values across three settings:
gasoline-engine highway driving, gasoline-engine urban driving, and
electric-vehicle highway driving. Performance is assessed using Top-1 and
Top-2 accuracy, macro F1, precision, recall, MCC, and the calibration metrics
Brier score and ECE.

\subsection{Single-Model Performance}
\label{subsec:single_model_results}

Table~\ref{tab:few_shot_big_observable_windows_average} summarizes the
single-model results under the few-shot setting. Mistral Small~24B provides the
strongest overall trade-off, achieving the highest Top-1 accuracy of 0.903, the
highest MCC of 0.887, and the lowest Brier score of 0.102. It is therefore
selected as the strongest single-model baseline.

Qwen2.5~32B remains highly competitive, with the highest macro F1 score of
0.895 and a Top-2 accuracy of 0.972, shared with Phi-4~14B and Gemma~3~27B.
Phi-4~14B also performs strongly, while the quantized Llama~3.1~70B does not
outperform the smaller medium-scale models. This suggests that, in this
structured diagnostic setting, model size alone is less important than the
ability to use the compact evidence representation effectively.

\begin{table}[h]
  \centering
  \caption{Average few-shot single-model performance across the evaluated
  scenarios. Bold values indicate the best value in each metric.}
  \label{tab:few_shot_big_observable_windows_average}
  \setlength{\tabcolsep}{2pt}
  \renewcommand{\arraystretch}{1.08}
  \begin{tabular}{@{}lcccccc@{}}
    \toprule
    Model & Top-1 & Top-2 & F1 & MCC & Brier $\downarrow$ & ECE $\downarrow$ \\
    \midrule
    Qwen2.5~14B
      & 0.877 & 0.927 & 0.872 & 0.841 & 0.152 & 0.201 \\
    Phi-4~14B
      & 0.896 & \textbf{0.972} & 0.894 & 0.875 & 0.129 & 0.217 \\
    \textbf{Mistral~24B}
      & \textbf{0.903} & 0.951 & 0.894 & \textbf{0.887} & \textbf{0.102} & 0.182 \\
    Gemma~3~27B
      & 0.829 & \textbf{0.972} & 0.830 & 0.757 & 0.144 & \textbf{0.114} \\
    Qwen2.5~32B
      & 0.897 & \textbf{0.972} & \textbf{0.895} & 0.874 & 0.110 & 0.166 \\
    DeepSeek-Qwen~32B
      & 0.859 & 0.965 & 0.857 & 0.810 & 0.134 & 0.153 \\
    Llama~3.1~70B
      & 0.787 & 0.965 & 0.784 & 0.737 & 0.168 & 0.218 \\
    \bottomrule
  \end{tabular}

  \medskip
  \footnotesize
  Mistral~24B denotes Mistral Small~24B. DeepSeek-Qwen~32B denotes DeepSeek R1
  Distill Qwen~32B. Llama~3.1~70B denotes the AWQ INT4 variant.
\end{table}

\subsection{Ensemble of LLMs Performance}
\label{subsec:top3_ensemble_results}

After evaluating the individual LLMs, the three strongest and most
complementary models were selected for ensemble fusion: Mistral Small~24B,
Qwen2.5~32B, and Phi-4~14B. These models were chosen because they achieved the
strongest single-model results in terms of Top-1 accuracy, macro F1, MCC, and
Top-2 ranking capability. Three ensemble strategies were then compared:
confidence-weighted voting, Borda rank aggregation, and reciprocal rank fusion.

Table~\ref{tab:top3_ensemble_by_scenario} reports the Top-3 ensemble results
for each evaluated use case. For the gasoline-engine highway case, all three
fusion strategies achieve the same Top-1 accuracy of 0.750, while Borda rank
aggregation and reciprocal rank fusion improve Top-2 accuracy to 0.917. This
indicates that rank-based fusion helps retain the correct class among the two
most likely candidates in the most difficult use case. However,
confidence-weighted voting remains more stable in terms of calibration. In the
gasoline-engine urban and electric-vehicle highway cases, confidence-weighted
voting achieves perfect classification performance, with 1.000 Top-1 accuracy,
F1, recall, precision, and MCC. It also provides the lowest Brier and ECE values
in these two use cases, showing that its confidence estimates are more reliable
than those of the rank-based fusion methods.

\begin{table}[t]
  \centering
  \caption{Performance of Top-3 ensemble strategies across the evaluated use
  cases. Bold values indicate the best value within each use case.}
  \label{tab:top3_ensemble_by_scenario}
  \setlength{\tabcolsep}{2pt}
  \renewcommand{\arraystretch}{1.08}
  \begin{tabular}{@{}llcccccc@{}}
    \toprule
    Case & Method & Top-1 & Top-2 & F1 & MCC & Brier $\downarrow$ & ECE $\downarrow$ \\
    \midrule
    Gas-Hwy
      & CW & 0.750 & 0.875 & 0.740 & 0.705 & 0.450 & 0.160 \\
      & \textbf{Borda}
      & \textbf{0.750} & \textbf{0.917} & 0.740 & \textbf{0.705} & \textbf{0.440} & \textbf{0.127} \\
      & RRF
      & 0.750 & \textbf{0.917} & \textbf{0.744} & 0.702 & 0.558 & 0.257 \\
    \midrule
    Gas-Urb
      & \textbf{CW}
      & \textbf{1.000} & \textbf{1.000} & \textbf{1.000} & \textbf{1.000} & \textbf{0.000} & \textbf{0.000} \\
      & Borda
      & \textbf{1.000} & \textbf{1.000} & \textbf{1.000} & \textbf{1.000} & 0.222 & 0.333 \\
      & RRF
      & \textbf{1.000} & \textbf{1.000} & \textbf{1.000} & \textbf{1.000} & 0.492 & 0.496 \\
    \midrule
    EV-Hwy
      & \textbf{CW}
      & \textbf{1.000} & \textbf{1.000} & \textbf{1.000} & \textbf{1.000} & \textbf{0.010} & \textbf{0.014} \\
      & Borda
      & \textbf{1.000} & \textbf{1.000} & \textbf{1.000} & \textbf{1.000} & 0.225 & 0.338 \\
      & RRF
      & \textbf{1.000} & \textbf{1.000} & \textbf{1.000} & \textbf{1.000} & 0.484 & 0.496 \\
    \bottomrule
  \end{tabular}

  \medskip
  \footnotesize
  Gas-Hwy: gasoline-engine highway; Gas-Urb: gasoline-engine urban;
  EV-Hwy: electric-vehicle highway; CW: confidence-weighted vote;
  RRF: reciprocal rank fusion.
\end{table}

\begin{table}[t]
  \centering
  \caption{Scenario-averaged performance of Top-3 ensemble strategies.}
  \label{tab:top3_ensemble_average}
  \setlength{\tabcolsep}{3pt}
  \begin{tabular}{@{}lcccccc@{}}
    \toprule
    Method & Top-1 & Top-2 & F1 & MCC & Brier $\downarrow$ & ECE $\downarrow$ \\
    \midrule
    \textbf{CW vote}
      & \textbf{0.917} & 0.958 & 0.913 & \textbf{0.902} & \textbf{0.153} & \textbf{0.058} \\
    Borda
      & \textbf{0.917} & \textbf{0.972} & 0.913 & \textbf{0.902} & 0.296 & 0.266 \\
    RRF
      & \textbf{0.917} & \textbf{0.972} & \textbf{0.915} & 0.901 & 0.511 & 0.416 \\
    \bottomrule
  \end{tabular}

  \vspace{2pt}
  \footnotesize
  CW vote: confidence-weighted vote; RRF: reciprocal rank fusion.
  The dagger marks the final selected ensemble strategy.
\end{table}

\begin{figure}[h]
  \centering
  \includegraphics[width=\linewidth]{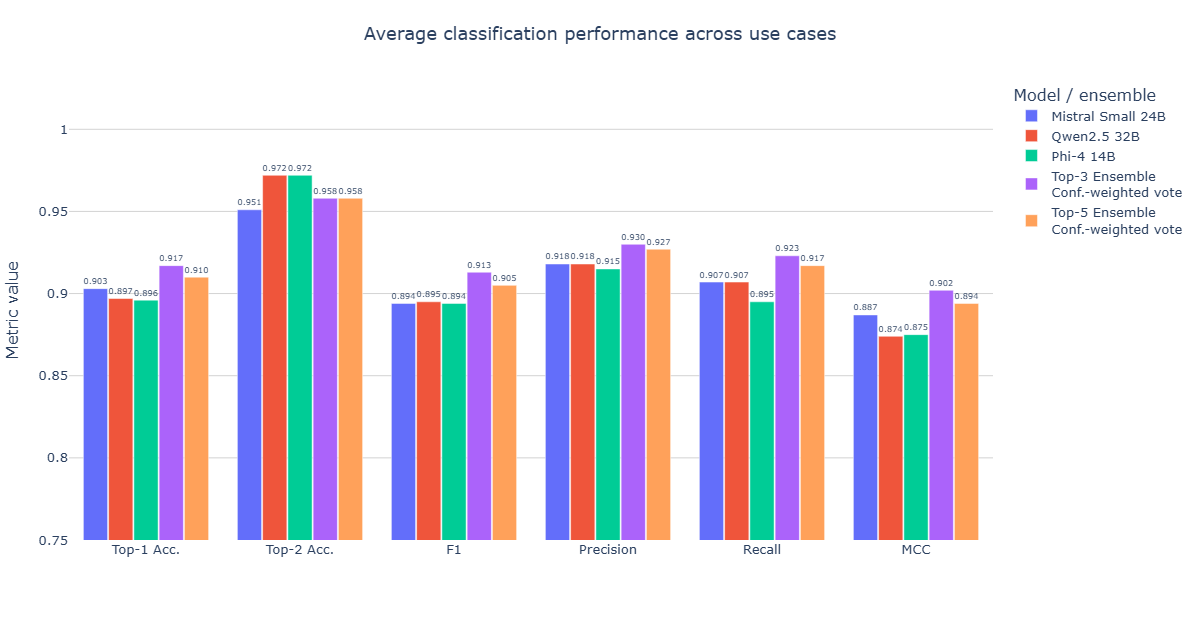}
  \caption{Scenario-averaged classification performance of the strongest
  single LLMs and the confidence-weighted Top-3 and Top-5 ensembles.}
  \Description{Bar chart comparing the scenario-averaged classification
  performance of the strongest individual LLMs with the Top-3 and Top-5
  confidence-weighted ensemble configurations.}
  \label{fig:single_vs_ensemble_classification}
\end{figure}

Figure~\ref{fig:single_vs_ensemble_classification} compares the three strongest
single models with the Top-3 confidence-weighted ensemble across the main
classification metrics. The Top-3 ensemble achieves the strongest overall
classification performance, with the best Top-1 accuracy, F1, recall, and MCC.
This indicates that combining complementary LLM predictions improves the
robustness of the final decision beyond the individual model outputs. The
single models remain competitive in Top-2 accuracy, suggesting that they often
rank the correct fault among the most likely candidates even when the Top-1
prediction is less reliable. Overall, the results support the use of the Top-3
ensemble as the final configuration, since it improves the main decision
quality metrics while preserving strong ranking performance.

The scenario-averaged results in
Table~\ref{tab:top3_ensemble_average} confirm that the Top-3 ensemble improves
the overall diagnostic performance. Confidence-weighted voting achieves 0.917
Top-1 accuracy, 0.913 macro F1, and 0.902 MCC. Although Borda rank aggregation
and reciprocal rank fusion reach a highered voting achieves 0.917
Top-1 accuracy, 0.913 macro F1, and  Top-2 accuracy of 0.972, their Brier
score and ECE are considerably worse. This suggests that rank-based fusion is
useful when the goal is to keep the correct class among the top candidates, but
it produces less reliable confidence scores. Therefore, confidence-weighted
voting is selected as the final Top-3 ensemble strategy because it provides the
best balance between classification accuracy, class-balanced reliability, and
calibration.

\begin{figure}[t]
\centering
\includegraphics[width=\linewidth]{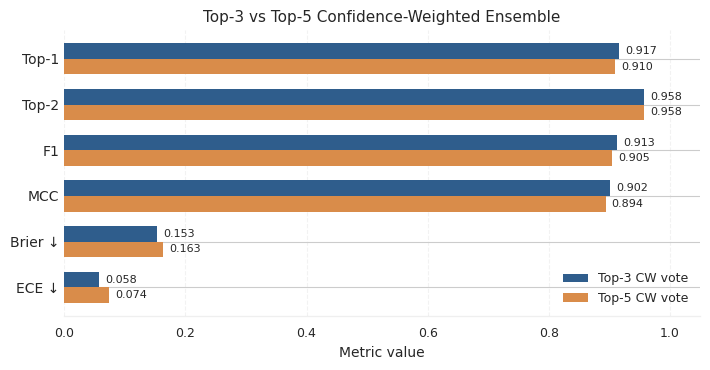}
\caption{Comparison of Top-3 and Top-5 confidence-weighted ensemble performance across classification and calibration metrics. Lower values are better for Brier and ECE.}
\label{fig:top3_vs_top5_all_metrics}
\end{figure}

For completeness, a Top-5 ensemble was also evaluated by adding Qwen2.5~14B and DeepSeek R1 Distill Qwen~32B to the Top-3 model set. Figure~\ref{fig:top3_vs_top5_all_metrics} compares the selected Top-3 confidence-weighted ensemble with the corresponding Top-5 variant. The Top-5 ensemble achieved 0.910 Top-1 accuracy, 0.905 macro F1, and 0.894 MCC, which are slightly below the Top-3 results. Top-2 accuracy remained unchanged, while the Brier score and ECE increased, indicating weaker confidence calibration. Thus, adding more models did not improve the ensemble in this setting. The result suggests that ensemble performance depends more on the complementarity and reliability of the selected models than on the number of models alone.

\begin{table}[t]
  \centering
  \caption{Representative explanation examples from Mistral Small~24B under the
  few-shot observable-window setting.}
  \label{tab:explanations}
  \setlength{\tabcolsep}{3pt}
  \renewcommand{\arraystretch}{1.08}
  \begin{tabular}{@{}p{0.29\columnwidth}p{0.66\columnwidth}@{}}
    \toprule
    \textbf{Field} & \textbf{Example 1} \\
    \midrule
    System & EV \\
    True / predicted & \texttt{BAT\_HV\_Gain\_3 / BAT\_HV\_Gain\_3} \\
    Fault location & High-voltage battery \\
    Confidence & 0.95 \\
    Explanation &
    Strong direct evidence from \texttt{V\_Bat\_HV[V]} with z=4.684 and
    md=6.671, supported by changes in \texttt{I\_Bat\_HV[A]},
    \texttt{T\_Bat\_HV[degC]}, and \texttt{SOC\_Bat\_HV[\%]}. \\
    \midrule
    \textbf{Field} & \textbf{Example 2} \\
    \midrule
    System & Gasoline engine \\
    True / predicted & \texttt{AccPedal\_Gain\_5 / AccPedal\_Gain\_5} \\
    Fault location & Accelerator pedal \\
    Confidence & 0.90 \\
    Explanation &
    Strong direct evidence for \texttt{AccPedal\_Gain\_5} from a significant
    change in \texttt{Pos\_AccPedal[\%]}. Medium-level changes in related
    support signals further reinforce the diagnosis. \\
    \bottomrule
  \end{tabular}
\end{table}
Beyond the predicted class and confidence score, the LLM outputs include a short explanation of the diagnostic evidence. Table~\ref{tab:explanations} shows representative examples from Mistral Small~24B under the few-shot observable-window setting. In both cases, the explanation refers first to the expected root-location signal, such as the high-voltage battery voltage for \texttt{BAT\_HV\_Gain\_3} or the accelerator-pedal position for \texttt{AccPedal\_Gain\_5}. Propagated signals are then used only as supporting evidence. This behavior is useful in HiL analysis because it helps the engineer check whether the model decision is based on a plausible root signal rather than only on downstream vehicle responses.

\subsection{Runtime and Model Complexity}
\label{subsec:runtime_complexity}

Table~\ref{tab:few_shot_big_observable_resource_complexity} reports the runtime
and memory characteristics of the evaluated LLMs. The fastest models are
Qwen2.5~14B, Phi-4~14B, and the quantized Llama~3.1~70B AWQ INT4, with
inference times below 3~s per window. However, the fastest models are not
necessarily the most accurate. Qwen2.5~32B achieves strong classification
performance but requires the longest inference time, with 6.614~s per window.
Mistral Small~24B represents a more balanced option, reaching the strongest
single-model diagnostic performance while requiring 4.940~s per window. GPU
memory usage is similar across models, at approximately 86~GB on the evaluation
hardware, whereas loading time varies considerably. Since loading is a one-time
cost, the per-window inference time is more relevant for practical offline
analysis. These results support using Mistral Small~24B as the strongest
single-model baseline, while the Top-3 ensemble is preferred when maximum
diagnostic performance is prioritized over latency.

\begin{table}[h]
  \centering
  \caption{Resource and complexity comparison of LLMs using the few-shot prompt.}
  \label{tab:few_shot_big_observable_resource_complexity}
  \setlength{\tabcolsep}{2pt}
  \renewcommand{\arraystretch}{1.08}
  \begin{tabular}{@{}lccccc@{}}
    \toprule
    Model & Inf. & Thr. & Out. & Load & Peak \\
          & (s)  &      & tok. & (s)  & (GB) \\
    \midrule
    Qwen14 & 2.747 & 0.364 & 99.681 & 100.031 & 86.296 \\
    Phi14 & 2.977 & 0.338 & 122.563 & 98.031 & 86.276 \\
    Mistral24 & 4.940 & 0.203 & 112.817 & 108.034 & 86.353 \\
    Gemma27 & 4.773 & 0.210 & 101.056 & 206.055 & 86.013 \\
    Qwen32 & 6.614 & 0.152 & 110.087 & 174.051 & 86.243 \\
    DS-Qwen32 & 5.879 & 0.171 & 92.515 & 82.025 & 86.271 \\
    Llama70-INT4 & 2.856 & 0.351 & 68.039 & 84.023 & 86.333 \\
    \bottomrule
  \end{tabular}

  \medskip
  \footnotesize
  Inf.: inference time per window; Thr.: throughput; Out. tok.: output tokens.
  Prompt tokens were constant at 1502.644 across models.
  DS-Qwen32 denotes DeepSeek R1 Distill Qwen~32B.
\end{table}

\section{Conclusion and future work}

\label{sec:conclusion}

This paper investigated open-weight LLM ensembles for fault-location classification in automotive HiL validation. The proposed pipeline avoids raw time-series prompting and instead compares faulty windows with time-aligned healthy references, converting the deviations into compact diagnostic evidence for few-shot prompting and ensemble fusion.

The results show that the Top-3 ensemble of Mistral Small~24B, Qwen2.5~32B, and Phi-4~14B gives the strongest overall performance across the evaluated gasoline-engine and electric-vehicle scenarios, reaching the best scenario-averaged Top-1 accuracy of 0.917. Confidence-weighted voting provided the best balance between classification accuracy and calibration. The Top-5 experiment further showed that adding more models does not necessarily improve the ensemble; in this case, the larger ensemble was slightly weaker, suggesting that model complementarity is more important than ensemble size alone. The generated explanations add a useful engineering layer by indicating which signals supported the prediction. This helps check whether the decision is based on plausible root-signal evidence rather than only on propagated vehicle responses.

Future work will extend the approach in several directions. First, fine-tuning or parameter-efficient adaptation should be investigated instead of relying only on prompting and in-context examples. This may improve consistency, calibration, and domain adaptation for HiL-specific signal patterns. Second, the method should be extended to concurrent faults, where multiple injected faults interact and the separation between root causes and propagated effects becomes more difficult. Further work should also evaluate larger datasets, additional driving profiles, stronger calibration strategies, and integration into continuous HiL test-bench workflows.

\bibliographystyle{ACM-Reference-Format}
\bibliography{sample-base}

\appendix

\end{document}